\documentstyle[12pt,emlines,bezier]{article}
\begin{document}
\title{
Demonstration of the quantum mechanics applicability limits
and solution of the irreversibility problem}

\author {V.A. Skrebnev \and R.N. Zaripov}
\maketitle

\hspace{1cm} Physics Department, University of Kazan, Kazan, Russia.

\begin{abstract}

An experimental study of the applicability of mechanics equations to describing
the process of equilibrium establishing in an isolated spin system was
performed. The time-reversion effects were used at the experiments.
It was demonstrated, that the equations of mechanics do not describe
the experimental results. The demonstration of the incomleteness of quantum 
mechanics description of the macrosystems annuls the contradiction which lays
in the basis of the irreversibility problem.
\end{abstract}

\section{Introduction}

     The problem of correlation between reversibility and irreversibility in
     the evolution of physical systems is one of the fundamental problems of
     contem\-po\-ra\-ry science. It is basically the problem  of the unsolved
     contradiction between the determined world that we get if we absolutize
     the mechanics equations , and the real world as we know it. At the level
     of the law of physics, this contradiction appears in the incompatibility
     of the 2nd law of thermodynamics and the reversibility of physical
     systems evolution followed from classical and quantum mechanics equations.

     In a numerous attempts to arrive to the irreversibility from the reversible
     equations there were an avowed or hidden assumptions that would allow the
     authors to arrive to a desirable result but would not have grounds within
     the mechanical theory.

     The reversibility of the evolution of those systems that are described
     by the mechanics equations has given birth to the idea that the source of
     the irreversibility is the interaction of the system with the surrounding
     environment where there are statistical laws in operation. The first
     thing that comes to mind when one reads the works proclaiming this point
     of view is the question of where those statistical laws in the physical
     objects surrounding the analyzed system come from.

      The irreversibility problem can neither be solved within the dynamical
chaos theory, since the chaoc in question is a determined and reversible one.

      The classical mechanics was so spectacularly successful and its
predictions were so exact, that it didn`t leave great Boltzmann the only and
the decisive argument in the discussion  with his opponents - he couldn`t have
made a statement about  the classical mechanics theory not being
complete.

Until early 20th centure there were no experimental data
demonstrating the incompleteness of the classical mechanics. The study of black
body radiation, atom spectra and photoeffect has lead to the creation of
quantum theory. But with all the empressive achievements of quantum mechanics,
there is our common sense and the fact, that the heat always goes from a hot
object to a cold one, which prevent us from deeming  the quantum mechanics
absolutely exact. This makes it crusual  to obtain experimental data which
would not be describable on the basis of quantum mechanics and would
demonstrate the manifestation of  the 2nd law of thermodynamics at the
evolution of macrosystems.

      It is practically impossible to solve the mechanics equations for the
macrosystems in which the thermodynamics laws apply. But if the mechanical
state of the system cannot be described, how can we compare  the
predictions of mechanics with those of thermodynamics? Nonetheless, such
comparison is possible, and it turns out that it does not require solving the
mechanics equations. This comparison can be made on the basis of real
experiments in changing the time sign in the equations which describe the
macrosystem behaviour.

\subsection{Possibility of time reversion experiments}

      It is possible to speak about a time-reversing experiment in principle
      based on the fact that changing the system Hamiltonian sign is
      equivalent to changing the time sign.
      This statement becomes obvious if we present the system wave function as

\begin{equation} \label{1}
\Psi (t) =e^{-i{\cal H}t} \Psi (0)
\end{equation}
or write down the Liouville equation solution for the system density matrix:
\begin{equation} \label{2}
   \sigma (t) = \exp (-i {\cal H}t) \sigma (0) \exp (i{\cal H}t)
\end{equation}

    We should bear in mind, though, that we are  interested in the internal
evolution of the system, which is determind by the interactions between all
its particles. It might seem impossible to change the sign of interactions in
macrosystems where the thermodynamics laws apply. Nevertheless, at least for spin
systems, it turned out to be possible not only to change the Hamiltonian sign,
but also to compare the forecasts of mechanics and the effects of the 2nd law
of thermodynamics under those conditions.

     The Hamiltonian of  a spin system placed in a strong external constant
magnetic field $H_0=\frac {\omega _0}{\gamma }$, in a frame rotating with a
frequency $\omega _0$ around the axis Z
reads $^{(1)}$:
\begin{equation} \label{3}
{\cal H} = {\cal H}_d'=\sum _{i<j} a_{ij}[I_{zi}I_{zj} - \frac14
(I_{+i}I_{-j}+I_{-i}I_{+j})]
\end{equation}

       If  an alternating field of a resonant frequence and an amplitude
$\omega _1 /\gamma $ is applied to the system, then in the rotating frame
we get:
\begin{equation} \label{4}
{\cal H}_R= \omega _1 I_x + {\cal H}_d'
\end{equation}

      Going into the tilted rotating frame by a unitary transformation
determined by the operator $\exp (-i \frac {\pi }{2} I_y)$ , we obtain:
\begin{eqnarray} \label{5}
&{\cal H}_{TR}= \omega _1 I_z -\frac12 {\cal H}_d' +\frac38 P,& \nonumber \\
&P={\cal H}_d^{(2)} +{\cal H}_d^{(-2)},\quad
{\cal H}_d^{(2)}={\cal H}_d^{(-2)*}=\sum \limits _{i<j} a_{ij}I_{+i}I_{+j}.&
\end{eqnarray}

      According to (2), we
     write down for the density matrix of the system with the
     Hamiltonian  (5):
\begin{equation} \label{6}
\sigma (t)=\exp \{-i(\omega _1 I_z -\frac12 {\cal H}_d' +\frac38 P)\}\sigma
(0)
\exp \{i(\omega _1 I_z -\frac12 {\cal H}_d' +\frac38 P)\}
\end{equation}

       In a strong alternating field $H_1$, when $\omega _1 \gg \omega _L$
       ($\omega _L = \gamma H_L$, where the local field
       $H_L=\{[Tr({\cal H}'_d)^2]/[Tr(M_z^2)]\}^{1/2}$
       $^{(1)}$),
       the non-secular operator 3/8$P$ in (6) can
be neglected in the first approximation.
Then, the internal evolution of the
system will be described by the Hamiltonian $-\frac 12 {\cal H}'_d$, while
the evolution in a rotating frame with no alternating field is described by
the Hamiltonian ${\cal H}'_d$.

      The transfer to a tilted rotating frame is a formal operation and cannot
actually influence the system evolution. At the same time, the influence of alternating field
$90 ^\circ _y$ -pulse
on the system is described by the same operator that describes the transfer to
a tilted frame. Thus, if we combine the sufficiently long application of a
strong alternating field to the system with the application of short pulse,
we can create a situation where the internal interactions sign in the system
Hamiltonian will be changed with an accuracy determined by the correlation
between $\omega _1$ and $\omega _L$

\subsection{                  Magic echo}

      The density matrix
\begin{equation} \label{7}
\sigma = 1 - \beta \omega _0 I_x
\end{equation}
describes the state of the of the spin system in a strong
external constant field after the $90^\circ _y$-pulse was applied to the
system. Here $\beta ^{-1}$ is the system temperature.

      The transverse component of magnetization $I_x(t)$ causes the free
      induction signal.

       The exciting experiments of Rhim, Pines and Waugh $^{(2)}$ have shown, that
the free induction signal restores in time which is much longer than the
spin-spin relaxation time $T_2$ ($T_2$ is defined as the time required
for the free induction signal decay under normal conditions). Before the experiments $^{(2)}$
this time was considered
as a thermodynamic relaxation time in spin systems.
The phenomenon observed in $^{(2)}$ was called "magic echo".

      The transverse component  of magnetization operator $I_x(t)$, to which the free
induction signal corresponds, is non-diagonal in the energy representation.
Consequently, the 2nd law of thermodynamics should not apply to the $I_x(t)$
evolution. Meanwhile, if the irreversibility of macrosystems evolution does
exist, and if we define the time of  equilibrium state establishing in the system as
the time of irreversible disappearance of the non-diagonal matrix elements in
the density matrix, we require the condition
\begin{equation} \label{8}
\frac {\partial \sigma}{\partial t} =-i[{\cal H},\sigma ]=0
\end{equation}
to be fulfilled in equilibrium, which means that the density matrix should
become diagonal in energy representation. The disappearance of the
non-diagonal matrix elements in the density matrix at the irreversible process
of establishing the equilibrium in the system should, logically, also be
irreversible.

The fact that the $I_x(t)$ evolution was reversible in
the experiment $^{(2)}$ made the authors come to a conclusion, that the spin
temperature concept should be treated cautiously.

      Thus, on the one hand, the spin temperature concept is confirmed by a
huge number of experimental data, and it is hard to doubt that the spin system
energy levels are populated according to the Boltzmann distribution. On the
other hand, the non-diagonal operator $I_x(t)$ in the system density matrix does
not irreversibly disappear during a time period which considerably surpasses
$T_2$. If  we assume that the spin temperature concept is correct, and that the
spin system evolution is really irreversible, then the experiments $^{(2)}$ can
lead us to a conclusion that  $T_2$ is not a thermodynamic relaxation time.

      If the $I_x(t)$ evolution in experiments $^{(2)}$ turned out to be
irreversible, the authors of $^{(2)}$ would not have had a basis for claiming
that the spin temperature concept should be treated cautiously. But on the
other hand the $I_x(t)$ evolution in that case would not have been describable
on the basis of mechanics approach. Such result would
have directly demonstrated the incompleteness of the quantum theory and the
significance of it would have been equivalent to this fact.

      The operator $I_x =\sum \limits _i I_{xi}$  has a simple structure and
is a sum of one-particle operators. The modern pulse NMR has the methods
$^{(1)}$ which allow to bring the spin system to a state described by the
density matrix
\begin{equation} \label{9}
\sigma = 1-\beta {\cal H}'_d
\end{equation}
with high value of the inverse temperature $\beta $.

When $\theta _y$-pulse is applied to
the spin system, the density matrix is transformed to $^{(1)}$:
\begin{equation} \label{10}
\sigma _R =1-\beta [\frac12(3\cos^2\theta -1){\cal H}'_d
+\frac 38 P\sin ^2\theta  -\frac 34 Q\sin \theta \cos \theta ]
\end{equation}
where
$$
Q=\sum _{i<j} a_{ij}[I_{zi}(I_{+j}+I_{-j})+I_{zj}(I_{+i}+I_{-i})].
$$
and $P$ had been determined by (5).

      Operators $P$ and $Q$ correspond to the interactions of all particles of the
spin system. It would be natural to expect that the evolution of those operators
would be different from the operator $I_x$ evolution. The paper $^{(3)}$
studied the behaviour of the
operator $P$ and $Q$ under the condition of the time sign change.
In energy representation, $Q$ has only non-diagonal matrix elements, but the energy reservoir corresponds
to the operator (3/8)$P$ at the time of
alternating field being applied. The experiments were carried out on the nuclei
spin system of $^{19}$F in a single crystal CaF$_2$. A dipole magic echo was discovered - the
restoration of the signal confined to the dipole interactions in the time that
considarably surpasses $T_2$. It turned out that the peculiarities in the
behaviour of the operator (3/8)$P$ to which the energy reservoir corresponds
cannot be explained on the basis of reversible equations of mechanics.

      The purpose of paper $^{(3)}$ was to find the dipole magic echo and to
compare the behaviours of the magic echo signal from $P$ and $Q$ operators. The measurements were
performed in one crystal orientation with respect to the external constant
magnetic field at a relatively small range of the alternating field amplitudes.

      Very important conclusions regarding the irreversibility can be made from
the results of $^{(3)}$. That is why the need for the continuation of the time
reversing experiments in spin systems is obvious. Besides, some aspects of
$^{(3)}$ require correction.

\section{Experiments, discussion}

      In this paper we observe and study the dipole magic echo in a wide range
of external alternating field values at various crystal orientations relative
to the constant magnetic field.

       The measurements are performed on the CaF$_2$ single crystal, which
is very convenient for this type of experiments. The avarage value of dipole-
dipole interactions in different orientations is proportional to local field
$\omega _L/ \gamma $, where $\omega _L$ is determined by
$\omega _L=(M_2/3)^{1/2}$  $^{(1)}$. Using the value of the second
moment $M_2$ of the NMR line for CaF$_2$, given by Abragam $^{(4)}$ we find in the case
of [111], [110] and [100] orientations
$\omega _L/ \gamma $= 0.88, 1.25, 2.01 G respectively.

The pulse sequences which is similar to those presented in $^{(3)}$ were used to achieve the
goals. The temperature was 300 K. The time of nuclei spin-lattice relaxation
was 8 s. Since the duration of the time reversing part of the applied pulse
sequences was not more than 1 ms, so the analized system may be considered
isolated from the lattice during this time.

\subsection{Evolution of the (3/8)$P$ dipole subsystem under the
        conditions of time reversion}

       We studied the operator $P$ evolution under the conditions of the time
reversion by pulse sequence 1 (Fig.1).

\vspace{2cm}
\unitlength=1.00mm
\special{em:linewidth 0.4pt}
\linethickness{0.4pt}
\begin{picture}(146.00,133.00)
\put(10.00,90.00){\vector(1,0){130.00}}
\put(140.00,90.00){\vector(0,0){0.00}}
\emline{10.00}{90.00}{1}{10.00}{130.00}{2}
\emline{100.00}{130.00}{3}{100.00}{90.00}{4}
\emline{120.00}{90.00}{5}{120.00}{110.00}{6}
\emline{80.00}{110.00}{7}{80.00}{90.00}{8}
\emline{60.00}{110.00}{9}{100.00}{110.00}{10}
\emline{10.00}{90.00}{11}{10.00}{85.00}{12}
\emline{30.00}{85.00}{13}{30.00}{90.00}{14}
\emline{60.00}{90.00}{15}{60.00}{85.00}{16}
\emline{80.00}{85.00}{17}{80.00}{90.00}{18}
\emline{100.00}{90.00}{19}{100.00}{85.00}{20}
\emline{120.00}{85.00}{21}{120.00}{90.00}{22}
\put(10.00,20.00){\vector(1,0){130.00}}
\emline{10.00}{60.00}{23}{10.00}{15.00}{24}
\emline{30.00}{15.00}{25}{30.00}{20.00}{26}
\emline{60.00}{15.00}{27}{60.00}{40.00}{28}
\emline{80.00}{40.00}{29}{80.00}{15.00}{30}
\emline{100.00}{15.00}{31}{100.00}{60.00}{32}
\emline{60.00}{40.00}{33}{100.00}{40.00}{34}
\bezier{116}(10.00,110.00)(20.00,105.00)(30.00,90.00)
\bezier{116}(10.00,40.00)(20.00,35.00)(30.00,20.00)
\bezier{184}(120.00,35.00)(124.00,50.00)(131.00,20.00)
\bezier{184}(131.00,90.00)(125.00,120.00)(120.00,105.00)
\put(19.00,130.00){\makebox(0,0)[cc]{$90^oy$}}
\put(30.00,106.00){\makebox(0,0)[cc]{$H_{1,-x}$}}
\put(107.00,130.00){\makebox(0,0)[cc]{$90^oy$}}
\put(70.00,101.00){\makebox(0,0)[cc]{$H_{1,x}$}}
\put(90.00,101.00){\makebox(0,0)[cc]{$H_{1,-x}$}}
\put(137.00,104.00){\makebox(0,0)[cc]{$<I_y>$}}
\put(146.00,90.00){\makebox(0,0)[cc]{t}}
\put(109.00,85.00){\makebox(0,0)[cc]{$t_{1/2}$}}
\put(19.00,85.00){\makebox(0,0)[cc]{2ms}}
\put(45.00,85.00){\makebox(0,0)[cc]{3ms}}
\put(70.00,85.00){\makebox(0,0)[cc]{$t_{1/2}$}}
\put(90.00,85.00){\makebox(0,0)[cc]{$t_{1/2}$}}
\put(20.00,60.00){\makebox(0,0)[cc]{$90^oy$}}
\put(107.00,60.00){\makebox(0,0)[cc]{$90^oy$}}
\put(137.00,42.00){\makebox(0,0)[cc]{$<I_y>$}}
\put(146.00,20.00){\makebox(0,0)[cc]{t}}
\put(71.00,30.00){\makebox(0,0)[cc]{$H_{1,x}$}}
\put(90.00,30.00){\makebox(0,0)[cc]{$H_{1,-x}$}}
\put(110.00,15.00){\makebox(0,0)[cc]{$t_{1/2}$}}
\put(90.00,15.00){\makebox(0,0)[cc]{$t_{1/2}$}}
\put(70.00,15.00){\makebox(0,0)[cc]{$t_{1/2}$}}
\put(45.00,15.00){\makebox(0,0)[cc]{3ms}}
\put(20.00,15.00){\makebox(0,0)[cc]{2ms}}
\put(30.00,35.00){\makebox(0,0)[cc]{$H_{1,-x}$}}
\emline{60.00}{110.00}{35}{60.00}{90.00}{36}
\bezier{184}(120.00,35.00)(116.00,50.00)(109.00,20.00)
\emline{120.00}{35.00}{37}{120.00}{20.00}{38}
\emline{120.00}{20.00}{39}{120.00}{20.00}{40}
\emline{120.00}{20.00}{41}{120.00}{20.00}{42}
\emline{120.00}{85.00}{43}{120.00}{130.00}{44}
\emline{120.00}{110.00}{45}{120.00}{90.00}{46}
\put(130.00,130.00){\makebox(0,0)[cc]{$45^oy$}}
\emline{60.00}{60.00}{47}{60.00}{40.00}{48}
\put(67.00,60.00){\makebox(0,0)[cc]{$45^oy$}}
\put(2.00,133.00){\makebox(0,0)[cc]{1.}}
\put(2.00,63.00){\makebox(0,0)[cc]{2.}}
\end{picture}

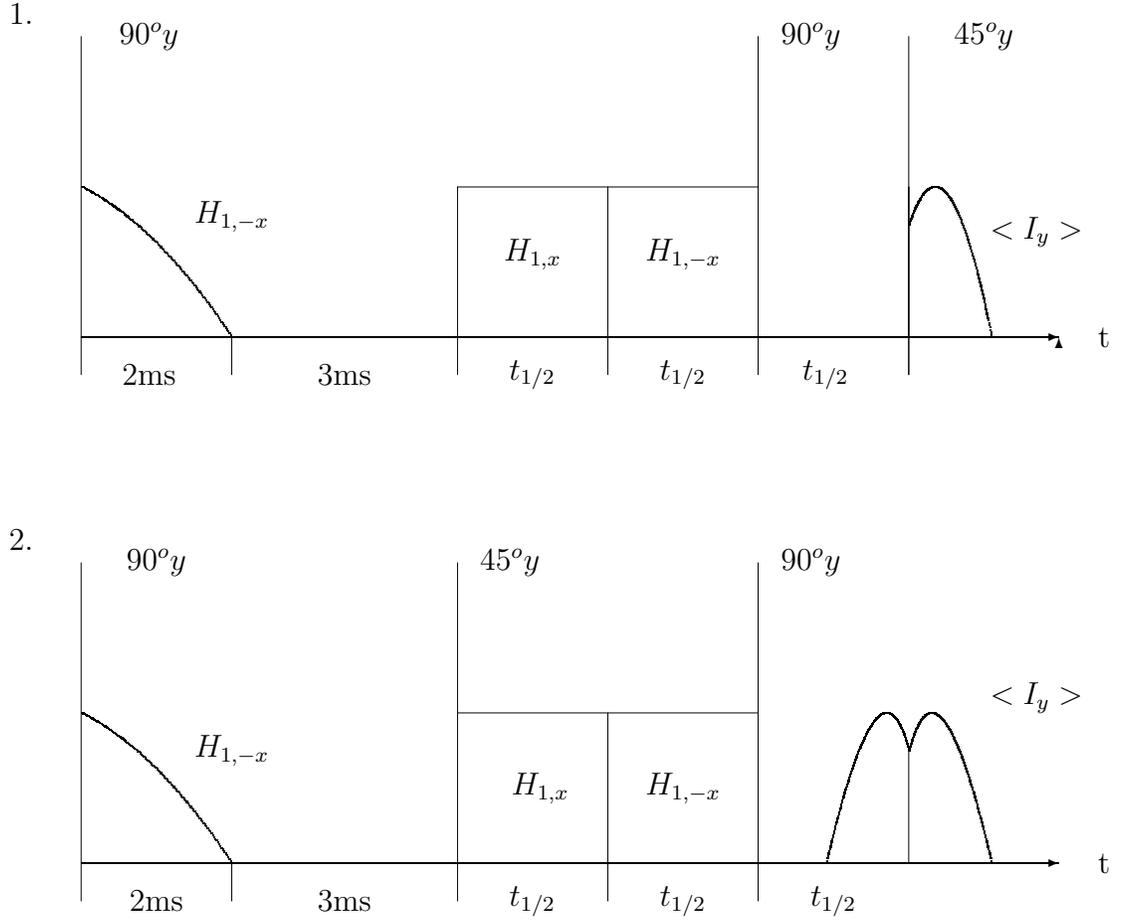
             Fig.1. Pulse sequences used
\vspace{1cm}

The alternating field phase was changed to opposite one in the
middle of the time interval of the application of this field.
       The adiobatic demagnetization in a rotating frame brings the system
       under consideration to a state which is described by the density
       matrix (9).
      The system Hamiltonian in a tilted rotating frame during the time
      of alternating field application is (5), and the initial density matrix
      is
\begin{equation}\label{11}
\sigma (0) =1-\beta (\frac 38 P -\frac 12 {\cal H}'_d)
\end{equation}

If the spin system obey the thermodynamic laws, it should be expected
that after the equilibrium is established in the system the density
matrix will look:
\begin{equation}\label{12}
\sigma _{eq}=1-\beta _1 (\omega _1 I_z +\frac 38 P-\frac 12 {\cal H}'_d)
\end{equation}

      The experiments  in  laboratory  $^{(1)}$  and  a   rotating
$^{(5)}$  frame  have  demonstrated,  that the unified spin system
temperature is established in two stages.  At the first stage, the
Zeeman  reservoir  and  the non-secular dipole-dipole interactions
reservoir (to which the operator (3/8)$P$ corresponds in our case)
undergo   the   quick  warm  mixing.  At  the  second  stage,  the
temperature of the newly created subsystem and of the reservoir of
the secular part of the dipole-dipole interactions get levelled at
a slower pace.

\subsubsection{Description of the operator $P$
evolution by reversible equations}

      Let's review the spin system evolution at the application of
the  pulse  sequence  1  on  the  basis  of  expression  (2).  The
transformation which corresponds to the 90$^\circ _y$-pulse effect
and to the transfer to the tilted rotating frame  compensate  each
other.
      That's why by the time the 45$^\circ $-pulse is applied we get:
\begin{eqnarray}\label{13}
&&\sigma (\frac 32 t_1)=A_1\sigma (0) A_1^{-1}\nonumber \\
&&A_1=\exp (-i{\cal H}'_d \frac {t_1}{2}) A_- A_+ \nonumber \\
&&A_{\pm }=\exp \{-i(\pm \omega _1 I_z +\frac 38 P-\frac 12 {\cal H}'_d)
\frac {t_1}{2}\}
\end{eqnarray}

      If the 45$^\circ $-pulse is excluded from the pulse sequence 1,
      then after the alternating field is removed the spin system signal
      is absent. The signal observed after the 45$^\circ $-pulse is
      determined by  the transverse component $I_y$, and we can write
      $\langle I_y\rangle  = \langle I_y\rangle _1 +\langle I_y\rangle _2$.
      The value $\langle I_y\rangle _2$  is the contribution to
      $\langle I_y \rangle $ which
      is bound to the density
      matrix operator $-\frac 12  {\cal  H}'_d$.  The  results  of
      $^{(5)}$  demonstrate,  that  this  contribution is constant
      under the conditions of our experiments and  can  be  easily
      subtracted from the signal under observation.

      If $\omega _1 \gg \omega _L$, then in the first approximation we can write
\begin{eqnarray}\label{14}
&&P(t_1 + t) = A_2 P A_2^{-1} \nonumber \\
&&A_2 =\exp \{-i{\cal H}'_d (t- \frac 12 t_1)\}
\end{eqnarray}

      It is demonstrated in $^{(3)}$, that in the case of
      $t_1=N\frac{\pi}{\omega _1}$  from (14) it follows:
\begin{equation} \label{15}
\langle I_y \rangle _1 =\frac{3}{16} \beta Tr (I_y^2)\frac{d}{dt} G(t)
\end{equation}
where $G(t)$ determines the form of the free induction signal.

      A signal whose amplitude does not depend upon the alternating field
      application time corresponds to (15).

Fig. 2 presents the dependence of the amplitude of the signal
corresponding to the (3/8)$P$ operator in the density matrix upon the
alternating field application time $t_1$.

      It follows from the Fig. 2
      that the signal in the pulse sequence 1 decays as $t_1$ grows.
      The signal decay turned out to grow with the transition from the
      [111] to [100] orientation,
      but not depend upon the $\omega _1$ in every orientation.

      If $t_1 =N\frac {\pi}{\omega _1}$, then, using the Magnus
      expansion $^{(6)}$, we find:
\begin{eqnarray}\label{16}
&&\exp \{-i(\omega _1 I_z +\frac 38 P-\frac 12 {\cal H}'_d)t\}=
\exp (-i\omega _1 I_z t)\exp (-iFt), \\
&&F=-\frac 12 {\cal H}'_d +
\sum ^\infty _{k=1} \frac{1}{(k+1)!}\frac {{\cal H}_k}
{(2\omega _1)^k}\nonumber
\end{eqnarray}

      The ${\cal H}_k$ operators have the dipole-dipole interactions
      magnitude in  the $k+1$ power. The correction to the average
      Hamiltonian $-\frac 12 {\cal H}'_d$,
      corresponding to the first term of the sum by $k$ in (16), is
\begin{eqnarray}\label{17}
&&{\cal H}^{(1)}={\cal H}^{(1)}_1+{\cal H}^{(1)}_2 \\
&&{\cal H}^{(1)}_1=(\frac 38)^2
\frac { [{\cal H}^{(-2)}_d,{\cal H}^{(2)}_d ] }{2\omega _1}\quad
{\cal H}^{(1)}_2=\frac 38 \frac 12 \frac {[{\cal H}'_d,{\cal H}^{(-2)}_d-
{\cal H}^{(2)}_d ]}{2\omega _1} \nonumber
\end{eqnarray}

We shall not need an explicit form of the corrections to
$-\frac 12 {\cal H}'_d$ of a higher order.

If the alternating field phase changes often, like in $^{(2)}$, the odd power
corrections to $-\frac 12 {\cal H}'_d$ in (16) vanish $^{(2,6)}$.
In our experiments, the correction ${\cal H}^{(1)}$ does not
disappear, but  the signal decay contribution due to ${\cal H}^{(1)}$
decreases because of the phase change of alternating field.

\vspace{2cm}

\setlength{\unitlength}{0.240900pt}
\ifx\plotpoint\undefined\newsavebox{\plotpoint}\fi
\sbox{\plotpoint}{\rule[-0.200pt]{0.400pt}{0.400pt}}%
\begin{picture}(1500,900)(0,0)
\font\gnuplot=cmr10 at 10pt
\gnuplot
\sbox{\plotpoint}{\rule[-0.200pt]{0.400pt}{0.400pt}}%
\put(220.0,113.0){\rule[-0.200pt]{292.934pt}{0.400pt}}
\put(220.0,113.0){\rule[-0.200pt]{4.818pt}{0.400pt}}
\put(198,113){\makebox(0,0)[r]{0}}
\put(1416.0,113.0){\rule[-0.200pt]{4.818pt}{0.400pt}}
\put(220.0,304.0){\rule[-0.200pt]{4.818pt}{0.400pt}}
\put(198,304){\makebox(0,0)[r]{0.25}}
\put(1416.0,304.0){\rule[-0.200pt]{4.818pt}{0.400pt}}
\put(220.0,495.0){\rule[-0.200pt]{4.818pt}{0.400pt}}
\put(198,495){\makebox(0,0)[r]{0.50}}
\put(1416.0,495.0){\rule[-0.200pt]{4.818pt}{0.400pt}}
\put(220.0,686.0){\rule[-0.200pt]{4.818pt}{0.400pt}}
\put(198,686){\makebox(0,0)[r]{0.75}}
\put(1416.0,686.0){\rule[-0.200pt]{4.818pt}{0.400pt}}
\put(220.0,877.0){\rule[-0.200pt]{4.818pt}{0.400pt}}
\put(198,877){\makebox(0,0)[r]{1}}
\put(1416.0,877.0){\rule[-0.200pt]{4.818pt}{0.400pt}}
\put(220.0,113.0){\rule[-0.200pt]{0.400pt}{4.818pt}}
\put(220,68){\makebox(0,0){50}}
\put(220.0,857.0){\rule[-0.200pt]{0.400pt}{4.818pt}}
\put(355.0,113.0){\rule[-0.200pt]{0.400pt}{4.818pt}}
\put(355,68){\makebox(0,0){100}}
\put(355.0,857.0){\rule[-0.200pt]{0.400pt}{4.818pt}}
\put(490.0,113.0){\rule[-0.200pt]{0.400pt}{4.818pt}}
\put(490,68){\makebox(0,0){150}}
\put(490.0,857.0){\rule[-0.200pt]{0.400pt}{4.818pt}}
\put(625.0,113.0){\rule[-0.200pt]{0.400pt}{4.818pt}}
\put(625,68){\makebox(0,0){200}}
\put(625.0,857.0){\rule[-0.200pt]{0.400pt}{4.818pt}}
\put(760.0,113.0){\rule[-0.200pt]{0.400pt}{4.818pt}}
\put(760,68){\makebox(0,0){250}}
\put(760.0,857.0){\rule[-0.200pt]{0.400pt}{4.818pt}}
\put(896.0,113.0){\rule[-0.200pt]{0.400pt}{4.818pt}}
\put(896,68){\makebox(0,0){300}}
\put(896.0,857.0){\rule[-0.200pt]{0.400pt}{4.818pt}}
\put(1031.0,113.0){\rule[-0.200pt]{0.400pt}{4.818pt}}
\put(1031,68){\makebox(0,0){350}}
\put(1031.0,857.0){\rule[-0.200pt]{0.400pt}{4.818pt}}
\put(1166.0,113.0){\rule[-0.200pt]{0.400pt}{4.818pt}}
\put(1166,68){\makebox(0,0){400}}
\put(1166.0,857.0){\rule[-0.200pt]{0.400pt}{4.818pt}}
\put(1301.0,113.0){\rule[-0.200pt]{0.400pt}{4.818pt}}
\put(1301,68){\makebox(0,0){450}}
\put(1301.0,857.0){\rule[-0.200pt]{0.400pt}{4.818pt}}
\put(1436.0,113.0){\rule[-0.200pt]{0.400pt}{4.818pt}}
\put(1436,68){\makebox(0,0){500}}
\put(1436.0,857.0){\rule[-0.200pt]{0.400pt}{4.818pt}}
\put(220.0,113.0){\rule[-0.200pt]{292.934pt}{0.400pt}}
\put(1436.0,113.0){\rule[-0.200pt]{0.400pt}{184.048pt}}
\put(220.0,877.0){\rule[-0.200pt]{292.934pt}{0.400pt}}
\put(45,570){\makebox(0,0){$\Large \frac{<I_y>_1 }{<I_y>_{1 \max}}$}}
\put(828,23){\makebox(0,0){ {${t_1,(\mu s)} $}}}
\put(220.0,113.0){\rule[-0.200pt]{0.400pt}{184.048pt}}
\put(328,877){\makebox(0,0){$\triangle$}}
\put(369,846){\makebox(0,0){$\triangle$}}
\put(409,831){\makebox(0,0){$\triangle$}}
\put(450,747){\makebox(0,0){$\triangle$}}
\put(490,785){\makebox(0,0){$\triangle$}}
\put(571,724){\makebox(0,0){$\triangle$}}
\put(612,655){\makebox(0,0){$\triangle$}}
\put(652,625){\makebox(0,0){$\triangle$}}
\put(693,587){\makebox(0,0){$\triangle$}}
\put(733,556){\makebox(0,0){$\triangle$}}
\put(774,503){\makebox(0,0){$\triangle$}}
\put(814,396){\makebox(0,0){$\triangle$}}
\put(855,312){\makebox(0,0){$\triangle$}}
\put(896,266){\makebox(0,0){$\triangle$}}
\put(936,205){\makebox(0,0){$\triangle$}}
\put(977,220){\makebox(0,0){$\triangle$}}
\put(1017,174){\makebox(0,0){$\triangle$}}
\put(1058,151){\makebox(0,0){$\triangle$}}
\put(1098,121){\makebox(0,0){$\triangle$}}
\put(1139,113){\makebox(0,0){$\triangle$}}
\put(1179,121){\makebox(0,0){$\triangle$}}
\put(328,854){\makebox(0,0){$\triangle$}}
\put(369,778){\makebox(0,0){$\triangle$}}
\put(409,793){\makebox(0,0){$\triangle$}}
\put(450,571){\makebox(0,0){$\triangle$}}
\put(490,633){\makebox(0,0){$\triangle$}}
\put(531,602){\makebox(0,0){$\triangle$}}
\put(571,495){\makebox(0,0){$\triangle$}}
\put(612,434){\makebox(0,0){$\triangle$}}
\put(652,342){\makebox(0,0){$\triangle$}}
\put(693,296){\makebox(0,0){$\triangle$}}
\put(733,251){\makebox(0,0){$\triangle$}}
\put(774,228){\makebox(0,0){$\triangle$}}
\put(814,174){\makebox(0,0){$\triangle$}}
\put(855,182){\makebox(0,0){$\triangle$}}
\put(896,128){\makebox(0,0){$\triangle$}}
\put(936,136){\makebox(0,0){$\triangle$}}
\put(328,839){\makebox(0,0){$\triangle$}}
\put(409,678){\makebox(0,0){$\triangle$}}
\put(490,327){\makebox(0,0){$\triangle$}}
\put(531,273){\makebox(0,0){$\triangle$}}
\put(571,205){\makebox(0,0){$\triangle$}}
\put(612,166){\makebox(0,0){$\triangle$}}
\put(652,151){\makebox(0,0){$\triangle$}}
\put(693,113){\makebox(0,0){$\triangle$}}
\put(342,862){\circle{18}}
\put(382,839){\circle{18}}
\put(423,839){\circle{18}}
\put(463,770){\circle{18}}
\put(504,717){\circle{18}}
\put(544,671){\circle{18}}
\put(585,686){\circle{18}}
\put(625,633){\circle{18}}
\put(666,564){\circle{18}}
\put(706,518){\circle{18}}
\put(747,480){\circle{18}}
\put(787,472){\circle{18}}
\put(828,396){\circle{18}}
\put(869,312){\circle{18}}
\put(909,251){\circle{18}}
\put(950,212){\circle{18}}
\put(990,159){\circle{18}}
\put(1031,174){\circle{18}}
\put(1071,136){\circle{18}}
\put(1112,151){\circle{18}}
\put(1152,121){\circle{18}}
\put(1193,113){\circle{18}}
\put(342,846){\circle{18}}
\put(382,801){\circle{18}}
\put(423,762){\circle{18}}
\put(463,717){\circle{18}}
\put(504,655){\circle{18}}
\put(544,571){\circle{18}}
\put(585,541){\circle{18}}
\put(625,449){\circle{18}}
\put(666,388){\circle{18}}
\put(706,304){\circle{18}}
\put(747,243){\circle{18}}
\put(787,159){\circle{18}}
\put(828,159){\circle{18}}
\put(869,151){\circle{18}}
\put(909,121){\circle{18}}
\put(950,113){\circle{18}}
\put(342,854){\circle{18}}
\put(382,762){\circle{18}}
\put(423,640){\circle{18}}
\put(463,457){\circle{18}}
\put(504,350){\circle{18}}
\put(544,228){\circle{18}}
\put(585,159){\circle{18}}
\put(625,174){\circle{18}}
\put(666,121){\circle{18}}
\put(355,869){\raisebox{-.8pt}{\makebox(0,0){$\Box$}}}
\put(396,816){\raisebox{-.8pt}{\makebox(0,0){$\Box$}}}
\put(436,839){\raisebox{-.8pt}{\makebox(0,0){$\Box$}}}
\put(477,770){\raisebox{-.8pt}{\makebox(0,0){$\Box$}}}
\put(517,717){\raisebox{-.8pt}{\makebox(0,0){$\Box$}}}
\put(558,694){\raisebox{-.8pt}{\makebox(0,0){$\Box$}}}
\put(598,633){\raisebox{-.8pt}{\makebox(0,0){$\Box$}}}
\put(639,602){\raisebox{-.8pt}{\makebox(0,0){$\Box$}}}
\put(679,571){\raisebox{-.8pt}{\makebox(0,0){$\Box$}}}
\put(720,541){\raisebox{-.8pt}{\makebox(0,0){$\Box$}}}
\put(760,457){\raisebox{-.8pt}{\makebox(0,0){$\Box$}}}
\put(801,426){\raisebox{-.8pt}{\makebox(0,0){$\Box$}}}
\put(842,373){\raisebox{-.8pt}{\makebox(0,0){$\Box$}}}
\put(882,281){\raisebox{-.8pt}{\makebox(0,0){$\Box$}}}
\put(923,273){\raisebox{-.8pt}{\makebox(0,0){$\Box$}}}
\put(963,205){\raisebox{-.8pt}{\makebox(0,0){$\Box$}}}
\put(1004,144){\raisebox{-.8pt}{\makebox(0,0){$\Box$}}}
\put(1044,144){\raisebox{-.8pt}{\makebox(0,0){$\Box$}}}
\put(1085,151){\raisebox{-.8pt}{\makebox(0,0){$\Box$}}}
\put(1125,136){\raisebox{-.8pt}{\makebox(0,0){$\Box$}}}
\put(1166,136){\raisebox{-.8pt}{\makebox(0,0){$\Box$}}}
\put(355,839){\raisebox{-.8pt}{\makebox(0,0){$\Box$}}}
\put(396,778){\raisebox{-.8pt}{\makebox(0,0){$\Box$}}}
\put(436,762){\raisebox{-.8pt}{\makebox(0,0){$\Box$}}}
\put(477,686){\raisebox{-.8pt}{\makebox(0,0){$\Box$}}}
\put(517,655){\raisebox{-.8pt}{\makebox(0,0){$\Box$}}}
\put(558,533){\raisebox{-.8pt}{\makebox(0,0){$\Box$}}}
\put(598,480){\raisebox{-.8pt}{\makebox(0,0){$\Box$}}}
\put(639,403){\raisebox{-.8pt}{\makebox(0,0){$\Box$}}}
\put(679,357){\raisebox{-.8pt}{\makebox(0,0){$\Box$}}}
\put(720,243){\raisebox{-.8pt}{\makebox(0,0){$\Box$}}}
\put(760,228){\raisebox{-.8pt}{\makebox(0,0){$\Box$}}}
\put(801,174){\raisebox{-.8pt}{\makebox(0,0){$\Box$}}}
\put(842,121){\raisebox{-.8pt}{\makebox(0,0){$\Box$}}}
\put(882,159){\raisebox{-.8pt}{\makebox(0,0){$\Box$}}}
\put(923,113){\raisebox{-.8pt}{\makebox(0,0){$\Box$}}}
\put(355,816){\raisebox{-.8pt}{\makebox(0,0){$\Box$}}}
\put(396,701){\raisebox{-.8pt}{\makebox(0,0){$\Box$}}}
\put(436,610){\raisebox{-.8pt}{\makebox(0,0){$\Box$}}}
\put(477,411){\raisebox{-.8pt}{\makebox(0,0){$\Box$}}}
\put(517,312){\raisebox{-.8pt}{\makebox(0,0){$\Box$}}}
\put(558,243){\raisebox{-.8pt}{\makebox(0,0){$\Box$}}}
\put(598,205){\raisebox{-.8pt}{\makebox(0,0){$\Box$}}}
\put(639,136){\raisebox{-.8pt}{\makebox(0,0){$\Box$}}}
\put(679,136){\raisebox{-.8pt}{\makebox(0,0){$\Box$}}}
\sbox{\plotpoint}{\rule[-0.500pt]{1.000pt}{1.000pt}}%
\put(301,877){\usebox{\plotpoint}}
\put(301.00,877.00){\usebox{\plotpoint}}
\put(339.94,863.03){\usebox{\plotpoint}}
\multiput(342,862)(35.353,-21.756){0}{\usebox{\plotpoint}}
\multiput(355,854)(36.042,-20.595){0}{\usebox{\plotpoint}}
\put(375.89,842.29){\usebox{\plotpoint}}
\put(402.02,810.82){\usebox{\plotpoint}}
\put(428.35,778.78){\usebox{\plotpoint}}
\put(458.73,750.63){\usebox{\plotpoint}}
\put(487.02,720.31){\usebox{\plotpoint}}
\multiput(490,717)(31.600,-26.918){0}{\usebox{\plotpoint}}
\multiput(517,694)(16.801,-37.959){2}{\usebox{\plotpoint}}
\put(557.63,621.39){\usebox{\plotpoint}}
\put(584.63,590.31){\usebox{\plotpoint}}
\put(608.44,556.31){\usebox{\plotpoint}}
\put(631.54,521.86){\usebox{\plotpoint}}
\multiput(652,487)(24.043,-33.839){2}{\usebox{\plotpoint}}
\put(705.51,422.49){\usebox{\plotpoint}}
\multiput(706,422)(29.901,-28.794){0}{\usebox{\plotpoint}}
\put(735.36,393.64){\usebox{\plotpoint}}
\put(764.71,364.29){\usebox{\plotpoint}}
\put(794.61,335.52){\usebox{\plotpoint}}
\put(826.39,308.82){\usebox{\plotpoint}}
\put(859.43,283.73){\usebox{\plotpoint}}
\put(893.38,259.85){\usebox{\plotpoint}}
\multiput(896,258)(36.287,-20.160){0}{\usebox{\plotpoint}}
\put(929.48,239.40){\usebox{\plotpoint}}
\put(965.52,218.80){\usebox{\plotpoint}}
\put(1001.62,198.32){\usebox{\plotpoint}}
\multiput(1004,197)(39.801,-11.793){0}{\usebox{\plotpoint}}
\put(1041.29,186.33){\usebox{\plotpoint}}
\put(1081.25,175.11){\usebox{\plotpoint}}
\multiput(1085,174)(39.801,-11.793){0}{\usebox{\plotpoint}}
\put(1121.38,164.96){\usebox{\plotpoint}}
\put(1162.53,159.51){\usebox{\plotpoint}}
\put(1166,159){\usebox{\plotpoint}}
\put(301,877){\usebox{\plotpoint}}
\put(301.00,877.00){\usebox{\plotpoint}}
\put(337.22,856.73){\usebox{\plotpoint}}
\multiput(342,854)(27.187,-31.369){0}{\usebox{\plotpoint}}
\put(366.03,827.19){\usebox{\plotpoint}}
\multiput(369,824)(20.426,-36.138){0}{\usebox{\plotpoint}}
\multiput(382,801)(21.013,-35.800){2}{\usebox{\plotpoint}}
\multiput(409,755)(12.703,-39.520){2}{\usebox{\plotpoint}}
\put(456.39,642.30){\usebox{\plotpoint}}
\put(485.92,613.48){\usebox{\plotpoint}}
\put(504.43,576.86){\usebox{\plotpoint}}
\multiput(517,548)(16.801,-37.959){2}{\usebox{\plotpoint}}
\put(559.26,465.52){\usebox{\plotpoint}}
\put(581.75,430.68){\usebox{\plotpoint}}
\put(603.45,395.33){\usebox{\plotpoint}}
\multiput(625,365)(24.043,-33.839){2}{\usebox{\plotpoint}}
\put(678.74,296.30){\usebox{\plotpoint}}
\multiput(679,296)(27.769,-30.855){0}{\usebox{\plotpoint}}
\put(706.50,265.43){\usebox{\plotpoint}}
\put(734.07,234.56){\usebox{\plotpoint}}
\put(772.35,218.51){\usebox{\plotpoint}}
\put(809.27,199.63){\usebox{\plotpoint}}
\multiput(814,197)(36.591,-19.602){0}{\usebox{\plotpoint}}
\put(846.16,180.77){\usebox{\plotpoint}}
\put(885.96,168.97){\usebox{\plotpoint}}
\multiput(896,166)(36.287,-20.160){0}{\usebox{\plotpoint}}
\put(923.13,150.93){\usebox{\plotpoint}}
\put(960.66,134.42){\usebox{\plotpoint}}
\put(1001.72,128.34){\usebox{\plotpoint}}
\multiput(1004,128)(41.063,-6.083){0}{\usebox{\plotpoint}}
\put(1042.84,122.68){\usebox{\plotpoint}}
\put(1058,121){\usebox{\plotpoint}}
\put(301,877){\usebox{\plotpoint}}
\put(301.00,877.00){\usebox{\plotpoint}}
\put(332.12,849.58){\usebox{\plotpoint}}
\multiput(342,839)(27.187,-31.369){0}{\usebox{\plotpoint}}
\multiput(355,824)(9.143,-40.492){2}{\usebox{\plotpoint}}
\put(377.63,736.79){\usebox{\plotpoint}}
\multiput(382,724)(9.567,-40.394){3}{\usebox{\plotpoint}}
\put(423.12,577.57){\usebox{\plotpoint}}
\multiput(436,548)(11.808,-39.796){3}{\usebox{\plotpoint}}
\put(480.39,422.23){\usebox{\plotpoint}}
\multiput(490,403)(15.319,-38.581){2}{\usebox{\plotpoint}}
\put(534.00,310.44){\usebox{\plotpoint}}
\put(557.87,276.48){\usebox{\plotpoint}}
\put(583.60,244.00){\usebox{\plotpoint}}
\put(609.38,211.56){\usebox{\plotpoint}}
\put(635.71,179.88){\usebox{\plotpoint}}
\put(669.58,156.23){\usebox{\plotpoint}}
\multiput(679,151)(40.183,-10.418){0}{\usebox{\plotpoint}}
\put(708.73,143.19){\usebox{\plotpoint}}
\put(749.02,133.63){\usebox{\plotpoint}}
\multiput(760,132)(41.063,-6.083){0}{\usebox{\plotpoint}}
\put(790.08,127.54){\usebox{\plotpoint}}
\put(831.23,122.15){\usebox{\plotpoint}}
\put(872.35,116.50){\usebox{\plotpoint}}
\put(896,113){\usebox{\plotpoint}}
\end{picture}

          Fig.2. Dependence on $t_1$ of the amplitude of
          the magic echo signal due to the operator $P$ at
          $ \frac{\omega_1}{\gamma} = 12.5 G (\Box);
          \frac{\omega_1}{\gamma} =  25.3  G (\circ);
          \frac{\omega_1}{\gamma} = 57.2 G (\triangle) $
           in the [100], [110] and [111] orientations

\vspace{1cm}

      Using (16), by the time the 45$^{\circ }$-pulse is applied we get
      (in the case of constant alternating field phase):
\begin{eqnarray}\label{18}
&&P(\frac 32 t_1)=A_3PA^{-1}_3 \\
&&A_3=T\exp [-i\int_{0}^{t_1} \exp (-\frac {i}{2} {\cal H}'_d t)
{\cal H}^{(1)}_1 \cdot \exp (\frac {i}{2} {\cal H}'_d t)dt] \nonumber
\end{eqnarray}

      Writing down (18) we neglected the terms with $k>1$ in the expression
      for $F$ (16), because under the conditions of our experiments the ratio
      of $-\frac 12 {\cal H}'_d$  to the
      $\frac {1}{(k+1)!}\frac {{\cal H}_k}{(2\omega _1)^k}$
      is of the order of 10$^4$ already for $k = 2$.
      Hence, the correction terms with $k>1$ cannot contribute considerably
      to the decay of the signal which is due to the operator (3/8)$P$.
      We ommitted also the non-secular operator ${\cal H}_2^{(1)}$ in
      expression for ${\cal H}^{(1)}$.

      Taking into account the change of the alternating field phase we have:
\begin{eqnarray}\label{19}
&&P(\frac 32 t_1)=A_4PA^{-1}_4 \\
&&A_4=\exp (-i {\cal H}'_d \frac {t_1}{2})
\exp [i(\frac{{\cal H}'_d}{2} +{\cal H}^{(1)}_1)\frac {t_1}{2}]
\exp [i(\frac{{\cal H}'_d}{2} -{\cal H}^{(1)}_1)\frac {t_1}{2}].
\nonumber
\end{eqnarray}

\subsubsection{Analysis of the alternating field ungomogeneity
influence on the experiments results}

      It was shown experimentally in $^{(3)}$, that the signal in the pulse
sequence 1 decays much faster in the case of the constant alternating field
phase. It would be natural to give the following explanation: when the
alternating field phase is constant, the contribution to the decay grows by
means of the operator ${\cal H}^{(1)}_1$. Besides, the changing of the
alternating field phase reverses the direction of the isochromates precession
in the rotating frame and
compensates the field inhomogeneity influence on the signal under observation.

      The time of the alternating field application in  both phases was
divisible by $\frac {\pi}{\omega _1}$. Fig. 2 demonstrates, that the signal
decay at pulse sequence 1 does not depend upon the alternating field
amplitude. At the same time, the
signal decay time turned out to vary at different crystal orientation.

       If it was operator ${\cal H}^{(1)}_1$ playing the major role in the
signal decay in
pulse sequence 1, then the signal would have been decaying at a slower pace
when the $\omega _1$ would grow. Also, if it was the not totally compensated field
inhomogeneity that determined the signal dependence on $t_1$,
the signal in pulse
sequence 1 would have been decreasing when $\omega _1$ would grow.
Since the field
inhomogeneity contribution to the signal decay is directly proportional
to $\omega _1$,
and the operator ${\cal H}^{(1)}_1$ contribution is in reverse proportion
to $\omega _1$, it is also
not possible to explain the signal independence from $\omega _1$ by sum of
those two contributions.

      We have also achieved purely experimental evidence of the fact that the
signal decay in pulse sequence 1 cannot be connected to the alternating field
inhomogeneity. Indeed, the alternating field inhomogeneity does not depend
upon the crystal orientation, and the fact that the signal heavily depends on
the orientation points out that the signal decay is not determined by the
field inhomogeneity even at its maximum. Finally, we have taken the
measurements where we used the coils of varying size to create the alternating
field, which, obviously, produced the fields of different homogeneity.
However, the measurement results did not change when the coils were changed.
It would also be mentioned that the measurements were performed in this work
and in $^{(3)}$ using different equipment and different samples.
The results for the
same orientations and with the same $\omega _1$ value coinsided,
which is the evidence of there correctness.

\subsubsection{ Irreversibility of the (3/8)$P$ subsystem evolution}

 The pulse sequence "c" used in $^{(3)}$ allow to measure the decay of (3/8)$P$
operator signal, which is due to the operator $-\frac 12 {\cal H}'_d$
in the Hamiltonian (5). This decay time occurs in orientation [111] was equal
to 120 $\mu $s.

 The operator ${\cal H}^{(1)}_1$, which, if the quantum
mechanics description is correct, determines the dipole magic echo signal decay,
is on the two orders of magnitude smaller than the operator  $-\frac 12 {\cal H}'_d$,
when the values of $\omega _1$ are those used in our experiments.
Correspondingly, even not taking the alternating field phase
change into account, the signal decay time $t_d$ in pulse sequence 1
should be of two orders longer than 120 $\mu $s. The alternating field
phase change reverses the sign of $\omega _1$ and this increases the expected
time $t_d$ even more (see (18) and (19)). Thus, there is ground to think, that the decay
times $t_d$ which are observed at
pulse sequence 1 and do not exceed 350 $\mu $s are much
shorter than the decay time that follows from the system description based on
expressions (18) and (19) which follows from (2).

      Next, at the transfer from one orientation to another, the value of the
operator ${\cal H}^{(1)}_1$ changes in proportion to the second power of the
local field, i.e. in proportion to the change of the second moment $M_2$, which
grows 5 times when the system transfered from orientation [111] to orientation
[100] $^{(4)}$ . Hence,
the difference between the value of the operator ${\cal H}^{(1)}_1$ in the
orientation [100]
when $\frac {\omega _1}{\gamma}= 12.5$G and in the orientation [111] when
$\frac {\omega _1}{\gamma}= 52.6$G is 20 times, and
the signal disappearence time should also be on 20 times different. Fig.2
demonstrates, though, that the times under comparison for the signal in
pulse sequence 1 are not more than 2 times different.

Finally, the signal in the pulse sequence 1 does not depend on
$\omega _1 $ contrary to the predictions of the theory based on equation (2).

      Thus, the behaviour of the dipole-dipole interaction subsystem, to which
the operator (3/8)$P$ corresponds, under the time reversion conditions cannot be
described on the basis of expressions (1) and (2). This
fact leads us to the following coupled conclusions: \\
a) the evolution of the subsystem corresponding to the operator (3/8)$P$ in
density matrix is irreversible;  \\
b) the process of a system coming to an equilibrium state really is irreversible and
cannot be described by the reversible expressions (1) and (2);\\
c) the macrosystem under analysis being isolated does not lead to the
reversibility of its evolution.

      At the same time, if the subsystem (3/8)$P$ evolution were reversible, it
would been the evidence either of the fact, that the spin temperature
concept is inapplicable, or of the fact that the system`s thermodynamic
relaxation time is much longer than the time of experiment, or of the fact that
the mechanics equations are fantastically exact and the thermodynamics
irreversibility is illusionary. But the 2nd law of thermodynamics is the
generalization of the experience and the discussion of its correctness makes no
sence, while the exactness of system evolution description on the basis of
mechanics equations is limited by the extent of the mechanical theory
completeness. The experimental results which we recieved studying the
evolution of   the   subsystem   of   non-secular    dipole-dipole
interactions, to which
the operator (3/8)$P$ corresponds, cannot be described by the reversible
mechanics equations and, hence, demonstrate the incompleteness of those
equations.

\subsubsection{Thermodynamical description of the evolution
of the (3/8)$P$ subsystem}

      We select the irreversible component of the evolution of the (3/8)$P$
      subsystem using pulse sequence 1. It makes sence to introduce its
temperature $\beta ^{-1}(t)$ for the thermodynamical
description of this subsystem
      irreversible evolution. The following integro-differential equation for
      the inverse temperature $\beta (t)$ was obtained in $^{(7)}$ on the basis of
      non-equilibrium \linebreak
      thermodynamics methods:
\begin{equation} \label{20}
d\beta /dt =-\int ^t_0 \beta (t') G_1(t'-t)dt',
\end{equation}
where
$$
     \begin{array}{rcl}

G_1(t'-t) & = & \frac{3^2}{8^2\langle {\cal H}_d^{(2)}
{\cal H}_d^{(-2)}\rangle } \sum _{i>j}
\langle [{\cal H}_{dij}^{(2)}, {\cal H}_d^{(-2)}]
\exp (i{\cal H}'_d (t'-t)/2)
\times   \\
 & & [{\cal H}_{dij}^{(-2)}, {\cal H}_d^{(2)}]
\exp (-i{\cal H}'_d (t'-t)/2)\rangle
\end{array}
$$

      It can be seen from (20), that the rate of the reverse temperature
change does not depend on $\omega _1$, which corresponds to the fact that
the signal observed in pulse sequence 1 does not depend on $\omega _1$.
      The $G_1(t)$ function cannot be calculated explicity.
      Let's write it down as
\begin{equation} \label{21}
G_1 (t'-t)=(n\omega _L)^2 G_2 (t'-t)
\end{equation}
where the value of $n$ is of the order of 1. Let us  use for $G_2(t)$ the
Gaussian approximation, which is natural for regular magnetic of the
CaF$_2$ type $^{(7)}$:
\begin{equation} \label{22}
G_2 (t'-t)=\exp (-\frac12 M(t'-t)^2).
\end{equation}

      The value  $M$ must be comparable to the second moment $M_2$
of the NMR line in CaF$_2$.  We get  $M_2  =  2.55  \cdot  10^{10}$,
$0.99\cdot 10^{10}$ and $0.5\cdot 10^{10}$ s$^{-2}$ for the [100],
[110] and [111] orientations $^{(4)}$.  Since operator $-\frac  12
{\cal H}'_d$ is in the function $G_1(t)$ exponent,  let's write $M
= M_2/4$. The dotted lines in the Fig.2 correspond to the equation
(20)  solution  when  n  =  0.45  and  when  $M = M_2/4$ for every
orientation.  The agreement with the experiment is the  best  when
the  starting  point  is  moved  80 $\mu $s in $t$ in the equation
(20).  It can be explained by the fact,  that when time  $t_1$  is
less  than  80  $\mu  $s,  the  system  stochastization  after the
alternating field application does not show yet. Anyhow, there was
no considerable signal decay at $t_1$ less than 80 $\mu $s.

      A total  coincidence  of  theory  and  experiment  when  the
relaxation  processes  are  described   by   the   non-equilibrium
thermodynamics  methods seems somewhat suspicious.  That is why we
decided not to try and select  such  version  of  the  correlation
function  $G_1(t)$  which  would  make  the  calculated  curves to
reproduce the experimental data exactly.  Besides,  the change  in
the  alternating  field phase can influence the subsystem (3/8)$P$
thermodynamics evolution and introduce  changes  in  the  equation
(20) description of this evolution,  especially at small values of
$t_1$.

   It can be seen from Fig.  2,  that  in  the  case  of  Gaussian
approximation  of  $G_2(t)$  function,  when  the parameter values
correspond to the sample  under  observation,  the  eguation  (20)
solution  describe  the  character  of  the  signal  dependence on
orientation and time very well.  The degree  of  the  quantitative
correspondence  of  the theory $^{(7)}$ to the measurement results
may be considered quite enough.

\subsection{Evolution of non-diagonal operator $Q$ under the
condition of time reversion}

   Let us now review the spin system evolution at pulse sequence 2
(Fig.1).
After the 45$^{\circ }$-pulse the system state is described by the
    density       matrix
\begin{equation} \label{23}
\sigma = 1- \frac 14 \beta ({\cal H}'_d +\frac 34 P - \frac 32 Q)
\end{equation}

      The operator $Q$ in (23) causes the signal $^{(1)}$:
\begin{equation} \label{24}
\langle I_y \rangle  =\frac{3}{8} \beta Tr (I_y^2)\frac{d}{dt} G(t)
\end{equation}

      Similar to (19)  we obtain:
\begin{equation} \label{25}
Q(\frac 32 t_1)=A_4 Q A_4^{-1}.
\end{equation}

      In accordance with (25),  when $t=\frac 32  t_1$,  a  signal
confined  with  the  operator  $Q$  appears without any additional
influence on the system.  The amplitude of this signal turned  out
to  depend  upon  both  the  orientation  of  the  sample  and the
alternating field amplitude.  Hence,  the measurement results  are
presented  on  Fig.3  -  Fig.5.  Figs.  3  and  4  demonstrate the
dependence of the dipole magic echo signal  upon  the  alternating
field application time in the orientation [100] and [110] when the
ratio $\frac {\omega _1}{\gamma }$ was  changed  in  the  interval
12.5 - 52.7 G. \vspace{2cm}

\setlength{\unitlength}{0.240900pt}
\ifx\plotpoint\undefined\newsavebox{\plotpoint}\fi
\sbox{\plotpoint}{\rule[-0.200pt]{0.400pt}{0.400pt}}%
\begin{picture}(1500,900)(0,0)
\font\gnuplot=cmr10 at 10pt
\gnuplot
\sbox{\plotpoint}{\rule[-0.200pt]{0.400pt}{0.400pt}}%
\put(220.0,113.0){\rule[-0.200pt]{292.934pt}{0.400pt}}
\put(220.0,113.0){\rule[-0.200pt]{4.818pt}{0.400pt}}
\put(198,113){\makebox(0,0)[r]{0}}
\put(1416.0,113.0){\rule[-0.200pt]{4.818pt}{0.400pt}}
\put(220.0,304.0){\rule[-0.200pt]{4.818pt}{0.400pt}}
\put(198,304){\makebox(0,0)[r]{0.25}}
\put(1416.0,304.0){\rule[-0.200pt]{4.818pt}{0.400pt}}
\put(220.0,495.0){\rule[-0.200pt]{4.818pt}{0.400pt}}
\put(198,495){\makebox(0,0)[r]{0.50}}
\put(1416.0,495.0){\rule[-0.200pt]{4.818pt}{0.400pt}}
\put(220.0,686.0){\rule[-0.200pt]{4.818pt}{0.400pt}}
\put(198,686){\makebox(0,0)[r]{0.75}}
\put(1416.0,686.0){\rule[-0.200pt]{4.818pt}{0.400pt}}
\put(220.0,877.0){\rule[-0.200pt]{4.818pt}{0.400pt}}
\put(198,877){\makebox(0,0)[r]{1}}
\put(1416.0,877.0){\rule[-0.200pt]{4.818pt}{0.400pt}}
\put(220.0,113.0){\rule[-0.200pt]{0.400pt}{4.818pt}}
\put(220,68){\makebox(0,0){50}}
\put(220.0,857.0){\rule[-0.200pt]{0.400pt}{4.818pt}}
\put(355.0,113.0){\rule[-0.200pt]{0.400pt}{4.818pt}}
\put(355,68){\makebox(0,0){100}}
\put(355.0,857.0){\rule[-0.200pt]{0.400pt}{4.818pt}}
\put(490.0,113.0){\rule[-0.200pt]{0.400pt}{4.818pt}}
\put(490,68){\makebox(0,0){150}}
\put(490.0,857.0){\rule[-0.200pt]{0.400pt}{4.818pt}}
\put(625.0,113.0){\rule[-0.200pt]{0.400pt}{4.818pt}}
\put(625,68){\makebox(0,0){200}}
\put(625.0,857.0){\rule[-0.200pt]{0.400pt}{4.818pt}}
\put(760.0,113.0){\rule[-0.200pt]{0.400pt}{4.818pt}}
\put(760,68){\makebox(0,0){250}}
\put(760.0,857.0){\rule[-0.200pt]{0.400pt}{4.818pt}}
\put(896.0,113.0){\rule[-0.200pt]{0.400pt}{4.818pt}}
\put(896,68){\makebox(0,0){300}}
\put(896.0,857.0){\rule[-0.200pt]{0.400pt}{4.818pt}}
\put(1031.0,113.0){\rule[-0.200pt]{0.400pt}{4.818pt}}
\put(1031,68){\makebox(0,0){350}}
\put(1031.0,857.0){\rule[-0.200pt]{0.400pt}{4.818pt}}
\put(1166.0,113.0){\rule[-0.200pt]{0.400pt}{4.818pt}}
\put(1166,68){\makebox(0,0){400}}
\put(1166.0,857.0){\rule[-0.200pt]{0.400pt}{4.818pt}}
\put(1301.0,113.0){\rule[-0.200pt]{0.400pt}{4.818pt}}
\put(1301,68){\makebox(0,0){450}}
\put(1301.0,857.0){\rule[-0.200pt]{0.400pt}{4.818pt}}
\put(1436.0,113.0){\rule[-0.200pt]{0.400pt}{4.818pt}}
\put(1436,68){\makebox(0,0){500}}
\put(1436.0,857.0){\rule[-0.200pt]{0.400pt}{4.818pt}}
\put(220.0,113.0){\rule[-0.200pt]{292.934pt}{0.400pt}}
\put(1436.0,113.0){\rule[-0.200pt]{0.400pt}{184.048pt}}
\put(220.0,877.0){\rule[-0.200pt]{292.934pt}{0.400pt}}
\put(45,570){\makebox(0,0){$\Large \frac{<I_y>}{<I_y>_{\max}}$}}
\put(828,23){\makebox(0,0){ {${t_1,(\mu s)} $}}}
\put(220.0,113.0){\rule[-0.200pt]{0.400pt}{184.048pt}}
\put(328,862){\makebox(0,0){$\triangle$}}
\put(355,839){\makebox(0,0){$\triangle$}}
\put(382,793){\makebox(0,0){$\triangle$}}
\put(409,778){\makebox(0,0){$\triangle$}}
\put(436,724){\makebox(0,0){$\triangle$}}
\put(463,694){\makebox(0,0){$\triangle$}}
\put(490,633){\makebox(0,0){$\triangle$}}
\put(517,610){\makebox(0,0){$\triangle$}}
\put(544,571){\makebox(0,0){$\triangle$}}
\put(571,503){\makebox(0,0){$\triangle$}}
\put(598,472){\makebox(0,0){$\triangle$}}
\put(625,457){\makebox(0,0){$\triangle$}}
\put(652,388){\makebox(0,0){$\triangle$}}
\put(679,327){\makebox(0,0){$\triangle$}}
\put(706,312){\makebox(0,0){$\triangle$}}
\put(733,289){\makebox(0,0){$\triangle$}}
\put(760,243){\makebox(0,0){$\triangle$}}
\put(787,228){\makebox(0,0){$\triangle$}}
\put(814,182){\makebox(0,0){$\triangle$}}
\put(842,189){\makebox(0,0){$\triangle$}}
\put(869,144){\makebox(0,0){$\triangle$}}
\put(896,151){\makebox(0,0){$\triangle$}}
\put(923,136){\makebox(0,0){$\triangle$}}
\put(950,121){\makebox(0,0){$\triangle$}}
\put(328,846){\circle{18}}
\put(355,793){\circle{18}}
\put(382,762){\circle{18}}
\put(409,709){\circle{18}}
\put(436,648){\circle{18}}
\put(463,602){\circle{18}}
\put(490,541){\circle{18}}
\put(517,495){\circle{18}}
\put(544,419){\circle{18}}
\put(571,388){\circle{18}}
\put(598,312){\circle{18}}
\put(625,304){\circle{18}}
\put(652,258){\circle{18}}
\put(679,212){\circle{18}}
\put(706,174){\circle{18}}
\put(733,174){\circle{18}}
\put(760,144){\circle{18}}
\put(787,136){\circle{18}}
\put(814,151){\circle{18}}
\put(842,121){\circle{18}}
\put(328,824){\raisebox{-.8pt}{\makebox(0,0){$\Box$}}}
\put(355,755){\raisebox{-.8pt}{\makebox(0,0){$\Box$}}}
\put(382,640){\raisebox{-.8pt}{\makebox(0,0){$\Box$}}}
\put(409,594){\raisebox{-.8pt}{\makebox(0,0){$\Box$}}}
\put(436,533){\raisebox{-.8pt}{\makebox(0,0){$\Box$}}}
\put(463,434){\raisebox{-.8pt}{\makebox(0,0){$\Box$}}}
\put(490,373){\raisebox{-.8pt}{\makebox(0,0){$\Box$}}}
\put(517,342){\raisebox{-.8pt}{\makebox(0,0){$\Box$}}}
\put(544,281){\raisebox{-.8pt}{\makebox(0,0){$\Box$}}}
\put(571,220){\raisebox{-.8pt}{\makebox(0,0){$\Box$}}}
\put(598,197){\raisebox{-.8pt}{\makebox(0,0){$\Box$}}}
\put(625,166){\raisebox{-.8pt}{\makebox(0,0){$\Box$}}}
\put(652,136){\raisebox{-.8pt}{\makebox(0,0){$\Box$}}}
\put(679,144){\raisebox{-.8pt}{\makebox(0,0){$\Box$}}}
\put(706,121){\raisebox{-.8pt}{\makebox(0,0){$\Box$}}}
\end{picture}

             Fig.3. Dependence on $t_1$ of the amplitude of
             the magic echo signal due to the operator $Q$ at
             $\frac{\omega_1}{\gamma} = 12.5 G (\Box);
              \frac{\omega_1}{\gamma} = 25.3 G (\circ);
               \frac{\omega_1}{\gamma} = 52.7 G (\triangle)$
               in the[100] orientation
\vspace{2cm}

\setlength{\unitlength}{0.240900pt}
\ifx\plotpoint\undefined\newsavebox{\plotpoint}\fi
\sbox{\plotpoint}{\rule[-0.200pt]{0.400pt}{0.400pt}}%
\begin{picture}(1500,900)(0,0)
\font\gnuplot=cmr10 at 10pt
\gnuplot
\sbox{\plotpoint}{\rule[-0.200pt]{0.400pt}{0.400pt}}%
\put(220.0,113.0){\rule[-0.200pt]{292.934pt}{0.400pt}}
\put(220.0,113.0){\rule[-0.200pt]{4.818pt}{0.400pt}}
\put(198,113){\makebox(0,0)[r]{0}}
\put(1416.0,113.0){\rule[-0.200pt]{4.818pt}{0.400pt}}
\put(220.0,304.0){\rule[-0.200pt]{4.818pt}{0.400pt}}
\put(198,304){\makebox(0,0)[r]{0.25}}
\put(1416.0,304.0){\rule[-0.200pt]{4.818pt}{0.400pt}}
\put(220.0,495.0){\rule[-0.200pt]{4.818pt}{0.400pt}}
\put(198,495){\makebox(0,0)[r]{0.50}}
\put(1416.0,495.0){\rule[-0.200pt]{4.818pt}{0.400pt}}
\put(220.0,686.0){\rule[-0.200pt]{4.818pt}{0.400pt}}
\put(198,686){\makebox(0,0)[r]{0.75}}
\put(1416.0,686.0){\rule[-0.200pt]{4.818pt}{0.400pt}}
\put(220.0,877.0){\rule[-0.200pt]{4.818pt}{0.400pt}}
\put(198,877){\makebox(0,0)[r]{1}}
\put(1416.0,877.0){\rule[-0.200pt]{4.818pt}{0.400pt}}
\put(220.0,113.0){\rule[-0.200pt]{0.400pt}{4.818pt}}
\put(220,68){\makebox(0,0){50}}
\put(220.0,857.0){\rule[-0.200pt]{0.400pt}{4.818pt}}
\put(355.0,113.0){\rule[-0.200pt]{0.400pt}{4.818pt}}
\put(355,68){\makebox(0,0){100}}
\put(355.0,857.0){\rule[-0.200pt]{0.400pt}{4.818pt}}
\put(490.0,113.0){\rule[-0.200pt]{0.400pt}{4.818pt}}
\put(490,68){\makebox(0,0){150}}
\put(490.0,857.0){\rule[-0.200pt]{0.400pt}{4.818pt}}
\put(625.0,113.0){\rule[-0.200pt]{0.400pt}{4.818pt}}
\put(625,68){\makebox(0,0){200}}
\put(625.0,857.0){\rule[-0.200pt]{0.400pt}{4.818pt}}
\put(760.0,113.0){\rule[-0.200pt]{0.400pt}{4.818pt}}
\put(760,68){\makebox(0,0){250}}
\put(760.0,857.0){\rule[-0.200pt]{0.400pt}{4.818pt}}
\put(896.0,113.0){\rule[-0.200pt]{0.400pt}{4.818pt}}
\put(896,68){\makebox(0,0){300}}
\put(896.0,857.0){\rule[-0.200pt]{0.400pt}{4.818pt}}
\put(1031.0,113.0){\rule[-0.200pt]{0.400pt}{4.818pt}}
\put(1031,68){\makebox(0,0){350}}
\put(1031.0,857.0){\rule[-0.200pt]{0.400pt}{4.818pt}}
\put(1166.0,113.0){\rule[-0.200pt]{0.400pt}{4.818pt}}
\put(1166,68){\makebox(0,0){400}}
\put(1166.0,857.0){\rule[-0.200pt]{0.400pt}{4.818pt}}
\put(1301.0,113.0){\rule[-0.200pt]{0.400pt}{4.818pt}}
\put(1301,68){\makebox(0,0){450}}
\put(1301.0,857.0){\rule[-0.200pt]{0.400pt}{4.818pt}}
\put(1436.0,113.0){\rule[-0.200pt]{0.400pt}{4.818pt}}
\put(1436,68){\makebox(0,0){500}}
\put(1436.0,857.0){\rule[-0.200pt]{0.400pt}{4.818pt}}
\put(220.0,113.0){\rule[-0.200pt]{292.934pt}{0.400pt}}
\put(1436.0,113.0){\rule[-0.200pt]{0.400pt}{184.048pt}}
\put(220.0,877.0){\rule[-0.200pt]{292.934pt}{0.400pt}}
\put(45,570){\makebox(0,0){$\Large \frac{<I_y>}{<I_y>_{\max}}$}}
\put(828,23){\makebox(0,0){ {${t_1,(\mu s)} $}}}
\put(220.0,113.0){\rule[-0.200pt]{0.400pt}{184.048pt}}
\put(328,869){\makebox(0,0){$\triangle$}}
\put(355,869){\makebox(0,0){$\triangle$}}
\put(382,862){\makebox(0,0){$\triangle$}}
\put(409,846){\makebox(0,0){$\triangle$}}
\put(436,831){\makebox(0,0){$\triangle$}}
\put(463,793){\makebox(0,0){$\triangle$}}
\put(490,785){\makebox(0,0){$\triangle$}}
\put(517,770){\makebox(0,0){$\triangle$}}
\put(544,762){\makebox(0,0){$\triangle$}}
\put(571,709){\makebox(0,0){$\triangle$}}
\put(598,694){\makebox(0,0){$\triangle$}}
\put(625,686){\makebox(0,0){$\triangle$}}
\put(652,648){\makebox(0,0){$\triangle$}}
\put(679,625){\makebox(0,0){$\triangle$}}
\put(706,610){\makebox(0,0){$\triangle$}}
\put(733,571){\makebox(0,0){$\triangle$}}
\put(760,526){\makebox(0,0){$\triangle$}}
\put(787,518){\makebox(0,0){$\triangle$}}
\put(814,480){\makebox(0,0){$\triangle$}}
\put(842,457){\makebox(0,0){$\triangle$}}
\put(869,411){\makebox(0,0){$\triangle$}}
\put(896,396){\makebox(0,0){$\triangle$}}
\put(923,357){\makebox(0,0){$\triangle$}}
\put(950,327){\makebox(0,0){$\triangle$}}
\put(977,319){\makebox(0,0){$\triangle$}}
\put(1004,304){\makebox(0,0){$\triangle$}}
\put(1031,251){\makebox(0,0){$\triangle$}}
\put(1058,258){\makebox(0,0){$\triangle$}}
\put(1085,220){\makebox(0,0){$\triangle$}}
\put(1112,212){\makebox(0,0){$\triangle$}}
\put(1139,189){\makebox(0,0){$\triangle$}}
\put(1166,197){\makebox(0,0){$\triangle$}}
\put(1193,182){\makebox(0,0){$\triangle$}}
\put(1220,144){\makebox(0,0){$\triangle$}}
\put(1247,166){\makebox(0,0){$\triangle$}}
\put(1274,128){\makebox(0,0){$\triangle$}}
\put(1301,113){\makebox(0,0){$\triangle$}}
\put(328,862){\circle{18}}
\put(355,862){\circle{18}}
\put(382,854){\circle{18}}
\put(409,824){\circle{18}}
\put(436,801){\circle{18}}
\put(463,778){\circle{18}}
\put(490,747){\circle{18}}
\put(517,717){\circle{18}}
\put(544,671){\circle{18}}
\put(571,678){\circle{18}}
\put(598,617){\circle{18}}
\put(625,594){\circle{18}}
\put(652,579){\circle{18}}
\put(679,533){\circle{18}}
\put(706,526){\circle{18}}
\put(733,480){\circle{18}}
\put(760,426){\circle{18}}
\put(787,396){\circle{18}}
\put(814,403){\circle{18}}
\put(842,350){\circle{18}}
\put(869,319){\circle{18}}
\put(896,304){\circle{18}}
\put(923,281){\circle{18}}
\put(950,235){\circle{18}}
\put(977,228){\circle{18}}
\put(1004,220){\circle{18}}
\put(1031,182){\circle{18}}
\put(1058,174){\circle{18}}
\put(1085,166){\circle{18}}
\put(1112,136){\circle{18}}
\put(1139,151){\circle{18}}
\put(1166,128){\circle{18}}
\put(1193,121){\circle{18}}
\put(328,846){\raisebox{-.8pt}{\makebox(0,0){$\Box$}}}
\put(355,846){\raisebox{-.8pt}{\makebox(0,0){$\Box$}}}
\put(382,801){\raisebox{-.8pt}{\makebox(0,0){$\Box$}}}
\put(409,793){\raisebox{-.8pt}{\makebox(0,0){$\Box$}}}
\put(436,755){\raisebox{-.8pt}{\makebox(0,0){$\Box$}}}
\put(463,732){\raisebox{-.8pt}{\makebox(0,0){$\Box$}}}
\put(490,709){\raisebox{-.8pt}{\makebox(0,0){$\Box$}}}
\put(517,648){\raisebox{-.8pt}{\makebox(0,0){$\Box$}}}
\put(544,610){\raisebox{-.8pt}{\makebox(0,0){$\Box$}}}
\put(571,602){\raisebox{-.8pt}{\makebox(0,0){$\Box$}}}
\put(598,564){\raisebox{-.8pt}{\makebox(0,0){$\Box$}}}
\put(625,503){\raisebox{-.8pt}{\makebox(0,0){$\Box$}}}
\put(652,480){\raisebox{-.8pt}{\makebox(0,0){$\Box$}}}
\put(679,457){\raisebox{-.8pt}{\makebox(0,0){$\Box$}}}
\put(706,442){\raisebox{-.8pt}{\makebox(0,0){$\Box$}}}
\put(733,380){\raisebox{-.8pt}{\makebox(0,0){$\Box$}}}
\put(760,342){\raisebox{-.8pt}{\makebox(0,0){$\Box$}}}
\put(787,335){\raisebox{-.8pt}{\makebox(0,0){$\Box$}}}
\put(814,281){\raisebox{-.8pt}{\makebox(0,0){$\Box$}}}
\put(842,266){\raisebox{-.8pt}{\makebox(0,0){$\Box$}}}
\put(869,205){\raisebox{-.8pt}{\makebox(0,0){$\Box$}}}
\put(896,197){\raisebox{-.8pt}{\makebox(0,0){$\Box$}}}
\put(923,182){\raisebox{-.8pt}{\makebox(0,0){$\Box$}}}
\put(950,151){\raisebox{-.8pt}{\makebox(0,0){$\Box$}}}
\put(977,166){\raisebox{-.8pt}{\makebox(0,0){$\Box$}}}
\put(1004,128){\raisebox{-.8pt}{\makebox(0,0){$\Box$}}}
\put(1031,136){\raisebox{-.8pt}{\makebox(0,0){$\Box$}}}
\put(1058,128){\raisebox{-.8pt}{\makebox(0,0){$\Box$}}}
\end{picture}

           Fig.4. Dpendence on $t_1$ of the amplitude of the
           magic echo signal due to the operator $Q$ at
           $\frac{\omega_1}{\gamma} = 12.5 G (\Box);
           \frac{\omega_1}{\gamma}= 25.3 G (\circ);
           \frac{\omega_1}{\gamma} =52.7 G (\triangle)$
           in the [110] orientation
\vspace{2cm}

\setlength{\unitlength}{0.240900pt}
\ifx\plotpoint\undefined\newsavebox{\plotpoint}\fi
\sbox{\plotpoint}{\rule[-0.200pt]{0.400pt}{0.400pt}}%
\begin{picture}(1500,900)(0,0)
\font\gnuplot=cmr10 at 10pt
\gnuplot
\sbox{\plotpoint}{\rule[-0.200pt]{0.400pt}{0.400pt}}%
\put(220.0,113.0){\rule[-0.200pt]{292.934pt}{0.400pt}}
\put(220.0,113.0){\rule[-0.200pt]{4.818pt}{0.400pt}}
\put(198,113){\makebox(0,0)[r]{0}}
\put(1416.0,113.0){\rule[-0.200pt]{4.818pt}{0.400pt}}
\put(220.0,304.0){\rule[-0.200pt]{4.818pt}{0.400pt}}
\put(198,304){\makebox(0,0)[r]{0.25}}
\put(1416.0,304.0){\rule[-0.200pt]{4.818pt}{0.400pt}}
\put(220.0,495.0){\rule[-0.200pt]{4.818pt}{0.400pt}}
\put(198,495){\makebox(0,0)[r]{0.50}}
\put(1416.0,495.0){\rule[-0.200pt]{4.818pt}{0.400pt}}
\put(220.0,686.0){\rule[-0.200pt]{4.818pt}{0.400pt}}
\put(198,686){\makebox(0,0)[r]{0.75}}
\put(1416.0,686.0){\rule[-0.200pt]{4.818pt}{0.400pt}}
\put(220.0,877.0){\rule[-0.200pt]{4.818pt}{0.400pt}}
\put(198,877){\makebox(0,0)[r]{1}}
\put(1416.0,877.0){\rule[-0.200pt]{4.818pt}{0.400pt}}
\put(331.0,113.0){\rule[-0.200pt]{0.400pt}{4.818pt}}
\put(331,68){\makebox(0,0){100}}
\put(331.0,857.0){\rule[-0.200pt]{0.400pt}{4.818pt}}
\put(552.0,113.0){\rule[-0.200pt]{0.400pt}{4.818pt}}
\put(552,68){\makebox(0,0){200}}
\put(552.0,857.0){\rule[-0.200pt]{0.400pt}{4.818pt}}
\put(773.0,113.0){\rule[-0.200pt]{0.400pt}{4.818pt}}
\put(773,68){\makebox(0,0){300}}
\put(773.0,857.0){\rule[-0.200pt]{0.400pt}{4.818pt}}
\put(994.0,113.0){\rule[-0.200pt]{0.400pt}{4.818pt}}
\put(994,68){\makebox(0,0){400}}
\put(994.0,857.0){\rule[-0.200pt]{0.400pt}{4.818pt}}
\put(1215.0,113.0){\rule[-0.200pt]{0.400pt}{4.818pt}}
\put(1215,68){\makebox(0,0){500}}
\put(1215.0,857.0){\rule[-0.200pt]{0.400pt}{4.818pt}}
\put(1436.0,113.0){\rule[-0.200pt]{0.400pt}{4.818pt}}
\put(1436,68){\makebox(0,0){600}}
\put(1436.0,857.0){\rule[-0.200pt]{0.400pt}{4.818pt}}
\put(220.0,113.0){\rule[-0.200pt]{292.934pt}{0.400pt}}
\put(1436.0,113.0){\rule[-0.200pt]{0.400pt}{184.048pt}}
\put(220.0,877.0){\rule[-0.200pt]{292.934pt}{0.400pt}}
\put(45,570){\makebox(0,0){$\Large \frac{<I_y>}{<I_y>_{\max}}$}}
\put(828,23){\makebox(0,0){ {${t_1,(\mu s)} $}}}
\put(220.0,113.0){\rule[-0.200pt]{0.400pt}{184.048pt}}
\put(308,877){\makebox(0,0){$\triangle$}}
\put(331,869){\makebox(0,0){$\triangle$}}
\put(353,854){\makebox(0,0){$\triangle$}}
\put(375,862){\makebox(0,0){$\triangle$}}
\put(397,854){\makebox(0,0){$\triangle$}}
\put(419,854){\makebox(0,0){$\triangle$}}
\put(441,846){\makebox(0,0){$\triangle$}}
\put(463,846){\makebox(0,0){$\triangle$}}
\put(485,816){\makebox(0,0){$\triangle$}}
\put(507,824){\makebox(0,0){$\triangle$}}
\put(530,785){\makebox(0,0){$\triangle$}}
\put(552,778){\makebox(0,0){$\triangle$}}
\put(574,762){\makebox(0,0){$\triangle$}}
\put(596,739){\makebox(0,0){$\triangle$}}
\put(618,732){\makebox(0,0){$\triangle$}}
\put(640,701){\makebox(0,0){$\triangle$}}
\put(662,686){\makebox(0,0){$\triangle$}}
\put(684,655){\makebox(0,0){$\triangle$}}
\put(706,640){\makebox(0,0){$\triangle$}}
\put(729,633){\makebox(0,0){$\triangle$}}
\put(751,602){\makebox(0,0){$\triangle$}}
\put(773,594){\makebox(0,0){$\triangle$}}
\put(795,548){\makebox(0,0){$\triangle$}}
\put(817,526){\makebox(0,0){$\triangle$}}
\put(839,518){\makebox(0,0){$\triangle$}}
\put(861,495){\makebox(0,0){$\triangle$}}
\put(883,472){\makebox(0,0){$\triangle$}}
\put(905,464){\makebox(0,0){$\triangle$}}
\put(927,419){\makebox(0,0){$\triangle$}}
\put(950,419){\makebox(0,0){$\triangle$}}
\put(972,403){\makebox(0,0){$\triangle$}}
\put(994,380){\makebox(0,0){$\triangle$}}
\put(1016,365){\makebox(0,0){$\triangle$}}
\put(1038,319){\makebox(0,0){$\triangle$}}
\put(1060,319){\makebox(0,0){$\triangle$}}
\put(1082,289){\makebox(0,0){$\triangle$}}
\put(1104,281){\makebox(0,0){$\triangle$}}
\put(1126,243){\makebox(0,0){$\triangle$}}
\put(1149,235){\makebox(0,0){$\triangle$}}
\put(1171,220){\makebox(0,0){$\triangle$}}
\put(1193,205){\makebox(0,0){$\triangle$}}
\put(1215,182){\makebox(0,0){$\triangle$}}
\put(1237,189){\makebox(0,0){$\triangle$}}
\put(1259,174){\makebox(0,0){$\triangle$}}
\put(1281,144){\makebox(0,0){$\triangle$}}
\put(1303,136){\makebox(0,0){$\triangle$}}
\put(1325,144){\makebox(0,0){$\triangle$}}
\put(1348,121){\makebox(0,0){$\triangle$}}
\put(308,862){\circle{18}}
\put(331,854){\circle{18}}
\put(353,869){\circle{18}}
\put(375,846){\circle{18}}
\put(397,839){\circle{18}}
\put(419,801){\circle{18}}
\put(441,785){\circle{18}}
\put(463,770){\circle{18}}
\put(485,762){\circle{18}}
\put(507,701){\circle{18}}
\put(530,694){\circle{18}}
\put(552,694){\circle{18}}
\put(574,640){\circle{18}}
\put(596,625){\circle{18}}
\put(618,610){\circle{18}}
\put(640,556){\circle{18}}
\put(662,556){\circle{18}}
\put(684,533){\circle{18}}
\put(706,480){\circle{18}}
\put(729,457){\circle{18}}
\put(751,442){\circle{18}}
\put(773,396){\circle{18}}
\put(795,365){\circle{18}}
\put(817,319){\circle{18}}
\put(839,335){\circle{18}}
\put(861,304){\circle{18}}
\put(883,251){\circle{18}}
\put(905,258){\circle{18}}
\put(927,220){\circle{18}}
\put(950,212){\circle{18}}
\put(972,182){\circle{18}}
\put(994,197){\circle{18}}
\put(1016,174){\circle{18}}
\put(1038,136){\circle{18}}
\put(1060,166){\circle{18}}
\put(1082,128){\circle{18}}
\put(1104,121){\circle{18}}
\put(308,854){\raisebox{-.8pt}{\makebox(0,0){$\Box$}}}
\put(331,831){\raisebox{-.8pt}{\makebox(0,0){$\Box$}}}
\put(353,778){\raisebox{-.8pt}{\makebox(0,0){$\Box$}}}
\put(375,762){\raisebox{-.8pt}{\makebox(0,0){$\Box$}}}
\put(397,717){\raisebox{-.8pt}{\makebox(0,0){$\Box$}}}
\put(419,686){\raisebox{-.8pt}{\makebox(0,0){$\Box$}}}
\put(441,633){\raisebox{-.8pt}{\makebox(0,0){$\Box$}}}
\put(463,602){\raisebox{-.8pt}{\makebox(0,0){$\Box$}}}
\put(485,556){\raisebox{-.8pt}{\makebox(0,0){$\Box$}}}
\put(507,495){\raisebox{-.8pt}{\makebox(0,0){$\Box$}}}
\put(530,464){\raisebox{-.8pt}{\makebox(0,0){$\Box$}}}
\put(552,449){\raisebox{-.8pt}{\makebox(0,0){$\Box$}}}
\put(574,380){\raisebox{-.8pt}{\makebox(0,0){$\Box$}}}
\put(596,319){\raisebox{-.8pt}{\makebox(0,0){$\Box$}}}
\put(618,304){\raisebox{-.8pt}{\makebox(0,0){$\Box$}}}
\put(640,281){\raisebox{-.8pt}{\makebox(0,0){$\Box$}}}
\put(662,243){\raisebox{-.8pt}{\makebox(0,0){$\Box$}}}
\put(684,228){\raisebox{-.8pt}{\makebox(0,0){$\Box$}}}
\put(706,182){\raisebox{-.8pt}{\makebox(0,0){$\Box$}}}
\put(729,182){\raisebox{-.8pt}{\makebox(0,0){$\Box$}}}
\put(751,144){\raisebox{-.8pt}{\makebox(0,0){$\Box$}}}
\put(773,144){\raisebox{-.8pt}{\makebox(0,0){$\Box$}}}
\put(795,113){\raisebox{-.8pt}{\makebox(0,0){$\Box$}}}
\end{picture}

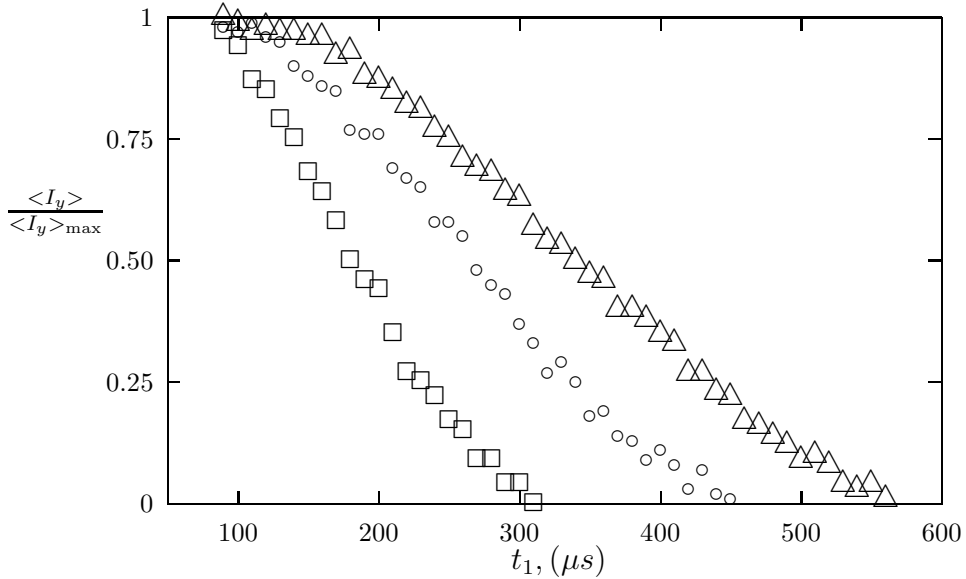
           Fig.5. Dependence on $t_1$ of the amplitude of the
           magic echo signal due to  the operator $Q$ at
           $\frac{\omega_1}{\gamma} = 52.7 G $ in the [100] - $(\Box)$;
           [110] - $(\circ)$; [111] - $(\triangle)$ orientation

\vspace{1cm}

Fig.5 demonstrates the signal dependence on $t_1$  with $\frac {\omega _1}{\gamma }$
 = 52.7 G at various crystal orientations.
 The measurement results for orientation [111] at $\frac {\omega _1}{\gamma }$=
12.5 G and $\frac {\omega _1}{\gamma }$ = 25.3 G coinside with those in
$^{(3)}$ and thus are not presented here.

      The facts that the signal in pulse sequence 2 grows when $\omega _1$
grows and
that the signal depends on orientation demonstrate that the alternating field
inhomogeneity influence on the effects under observation is neglectable small.
If we consider that it is the operator ${\cal H}^{(1)}_1$ that causes the decay of the signal
in pulse sequence 2, then the amplitude should grow when $P$ grows, and
decrease when $\omega _1$ grows. Figs. 3-5 show, that the experimental data
qualitatively
correspond to this conclusion. The inference that the operator $Q$ evolution
is reversible when a time-reversing pulse sequence is applied to the system was
made in $^{(3)}$, where the measurements were taken only in one orientation,
based on the signal dependence on $\omega _1$. However, in present work we have analized
the whole set of
experimental data that we obtained when studying the $Q$ magic echo in various
orientation and with a wide range of $\omega _1$ values,
and this analysis questions the correctness of above conclusion.

       Indeed, the signal decay time in pulse sequence 2, similar to the pulse sequence 1,
proves to be much less than that corresponding to presence
of the operator ${\cal H}^{(1)}_1$ in the expression for $A_4$.
Besides, the decay times in
orientation [111] at $\frac {\omega _1}{\gamma }$  = 52.7 G and in
orientation  [100] at $\frac {\omega _1}{\gamma }$= 12.5 G are
not more than 3 times different, while the value of the operator
${\cal H}^{(1)}_1$ on 20 times differs.
Thus, it is difficult to reconsile the idea proposed in $^{(3)}$ of the
operator $Q$ evolution
being reversible at pulse sequence 2 with the results of our experiments.

      The signal decay time $t_d$ in pulse sequence 2 is longer than that
in pulse sequence 1, but values of $t_d$ remain in the same order of magnitude.
According
to the basic statements of statistical physics, the evolution results for the
diagonal terms of the density matrix and the non-diagonal ones are totally
different. Hence, the characteristics of the $P$ and $Q$ operators evolution in the equilibrium
establishing process should also be different. In our experiments, this
difference appears in the fact that the signal corresponding to the operator $Q$
depends on $\omega _1$.

\subsection{Comparison of the peculiarities of free
induction signal magic echo and dipole magic echo}

   In ref.$^{(2)}$, the alternating field phase was changed every
$\frac {\pi}{\omega } $seconds when the
time reversion situation was created.
As a result, the corrections to the mean
Hamiltonian introduced by the odd powers of $\omega _1$ become a zero,
and the alternating field inhomogeneity gets compensated well.
The noticable free induction signal magic echo was observed up to
$t_1$ = 650 $\mu$s.
The authors in $^{(2)}$
see the reason for the signal decay in the influence of the alternating field
phase-changing pulses non-idealities. Actually, the alternating field
amplitude in $^{(2)}$ was 100 G, and 500 pulses correspond to 650 $\mu$s of
the alternating field application. At that pulse rate there really is ground to
assume that the magic echo signal  decay is explained by the pulse
non-idealities.

      We can not say how a very frequent alternating field phase change
influences the irreversible component of the system evolution. But the
possible demonstrations of the system evolution irreversibility  under the
physical conditions which are created by the phase changing field, can be
different from the case of the long alternating field application without a
phase change. It is therefore possible, that the time of the feasible
irreversibility demonstration when a phase-changing field influences the
system exeeds the signal decay time as a result of the
pulse non-idealities.

      In our experiments the pulse non-idealities could not have influenced
the observed signals considerably. But the time of the dipole magic echo
signal decay turned out to be of the same order of magnitude as the time of
the free induction signal magic echo decay. It gives us ground to assume, that
the evolution irreversibility of the transverse component of  magnetization $I_x$
can contribute to the signal decay in the experiment $^{(2)}$.

On the other hand,  in our  experiments  the energy redistribution
occurs in  spin  macrosystem  both  in  pulse sequence 1 and pulse
sequence 2.  This redistribution, in accordance  with  2nd  law  of
thermodynamics, is  irreversible  and  the  irreversible change of
density matrix diagonal terms corresponds to this redistribution.

The operators,  giving diagonal and  non-diagonal  density  matrix
terms, are  expressed  by means of the same one-particle operators
of nuclear moment components. Correspondingly, the irreversibility
of the  evolution of the density matrix diagonal terms may lead to
the irreversibility of the non-diagonal terms evolution also.

     At the same time, in conditions of the experiments $^{(2)}$,
the energy redistribution does not occur in spin macrosystem. As a
result,  the character of the operator $Q$ and $I_x$ evolution  in
time reversion experiments may be different.

We would like also to stress here,  that operators $P$ and $Q$ are
many-particle  ones  and  their  evolution  may  be  distinguished
essentially  from  the evolution of operator $I_x$ because of this
reason.

We see that the comparison of the paper $^{(2)}$ and our experiments
results give rise to many questions. The answers on this questions may be
given by new experiments only.

\section{Conclusion}

      The study of the dipole magic echo that we have performed in a wide
range of the values of $\omega _1$ and at various crystal orientations,
allow us to make the following main conclusions:

1) It turned out to be impossible to describe the evolution of the (3/8)$P$
subsystem in the time reversion situation on the basis of expressions (1) and
(2). The process of the system transfer to the equilibrium state, which is
described by the density matrix (12) in the tilted rotating frame, is
irreversible.

2) The dependence of the dipole magic echo signal confined with the non-diagonal
operator $Q$ upon $\omega _1$ and upon the value of the dipole-dipole interactions
corresponds qualitative to formula (25) which is derived from (2).
But the quantitative evaluations demonstrate, that the signal decay in pulse
sequence 2 happens much faster than it follows from the description based on
the reversible expresions (1) and (2).

      Probability assumptions are used in a apparent or hidden form when a
transfer from the reversible mechanics equations to the irreversible
non-equilibrium thermodynamic ones takes place. As a role, the introduction of
those assumptions is explained by the facts, that the mechanics equations
can not be solved exactly and that it is necessary to turn to the shortened
description of the non-equilibrium processes in the macroscopic systems. The
experimental and theoretical research performed in $^{(3)}$, $^{(7)}$ and in
this paper
demonstrate, that it is not the matter of the impossibility of solving the
reversible mechanics equations exactly, but the matter of their inapplicability
to the non-equilibrium processes in the macrosystems which obey the 2nd low of
thermodynamics. This circumstance will allow to take a new view at the problem
of chaos, both the classical and quantum ones.

The research that we have
performed by far does not exhaust the unique possibilities which the dipole
magic echo phenomenon presents for the study of the correlation between the
reversibility and irreversibility in the macrosystems evolution. We believe that the
continuation of the time reversing experiments in the spin systems will bring
a new and, possibly, unexpected results.

\section*{Acknowledgements}

      The authors are greatful to M. Ovchinnikov, L. Tagirov, A. Galeev and
S. Moiseev for the manifold help with
the work.

\vspace{2cm}
\centerline{ References}

\noindent 1. M. Goldman, Spin Teemperatureand Nuclear Magnetic Resonance in Solids
(Oxford: Clarendon, 1970).\\
2. W.-K. Rhim, A. Pines and J.S. Waugh, Phys. Rev. Lett. 25: 218-222 (1970),
Phys. Rev. B 3: 684-695 (1971).\\
3. V. A. Skrebnev  and V. A. Safin, J. Phys. C: Solid State Phys. 19:
4105-4114 (1986).\\
4. A. Abragam, The Principles of Nuclear Magnetism (Oxford: Clarendon, 1961).\\
5. V.A. Safin, V.A. Skrebnev and V.M. Vinokurov,  Zh. Eksp. Teor. Fiz.
87:1889-1893 (1984).\\
6.U. Haeberlen, High Resolution NMR in Solids (Academic Press, 1976)\\
7. V. A. Skrebnev,  J. Phys.: Condens. Matter 2: 2037-2044 (1989).
\end{document}